\documentclass[conference]{IEEEtran}
\usepackage{cite}
\usepackage{amsmath,amssymb,amsfonts}
\usepackage{algorithm}
\usepackage{algpseudocode}
\usepackage{graphicx}
\usepackage{textcomp}
\usepackage{optidef}
\usepackage{xcolor}
\usepackage{adjustbox}
\def\BibTeX{{\rm B\kern-.05em{\sc i\kern-.025em b}\kern-.08em
    T\kern-.1667em\lower.7ex\hbox{E}\kern-.125emX}}
\begin{document}

\title{Chaotic Quantum Behaved Particle Swarm Optimization for Multiobjective Optimization in Habitability Studies\\
}

\author{\IEEEauthorblockN{1\textsuperscript{st} Arun John}
\IEEEauthorblockA{\textit{Computer Science and Engineering} \\
\textit{PES University}\\
Bangalore, India \\
arunjoh@gmail.com}
\and
\IEEEauthorblockN{2\textsuperscript{nd} Anish Murthy}
\IEEEauthorblockA{\textit{Computer Science and Engineering} \\
\textit{PES University}\\
Bangalore, India \\
murthy.anish@gmail.com}
\\

}
\maketitle

\begin{abstract}
In this paper, based on the Quantum-behaved Particle Swarm Optimization algorithm in \cite{QPSO_MOD}\cite{QPSO-original}\cite{QPSO_book}, we evolve the algorithm to optimize a multiobjective optimization problem, namely the Cobb Douglas Habitability function which is based on CES production functions in Economics. We also propose some changes to the Quantum-behaved Particle Swarm Optimization algorithm to mitigate the problem of the algorithm prematurely converging and show the results of the proposed changes to the Quantum-behaved Particle Swarm Optimization.
\end{abstract}


\section{Introduction}
Quantum-behaved Particle Swarm Optimization (QPSO) algorithm, proposed by Jun Sun, is an evolution of the Particle Swarm Optimization originally proposed by Kennedy and Ebenhart in 1995. \\
Particle Swarm Optimization (PSO) is an evolutionary optimization technique, which simulates the knowledge evolvement of a social organism, in which the individuals representing the candidate solutions fly through the multidimensional space to find an optima or sub optima. These particles are characterised by a position and a velocity in the multidimensional space and evaluate their position to a goal (fitness function) in every iteration and particles in a local neighborhood share memories of their best positions and use them to adjust their own velocities.\\
PSO is a distributed method that requires simple mathematical operators and short segments of code, making it an optimal solution where computational resources are at a premium. Its implementation is highly parallelizable and scales with the dimensionality of the search space. The standard PSO algorithm does not deal with constraints but, through variations in initializing and updating particles, constraints are straightforward to represent and adhere to.\\
Quantum-behaved Particle Swarm Optimization (QPSO) is a quantum model of the original PSO where the state of a particle is depicted by a wave-function $\psi(\overrightarrow{x},t)$, instead of a position and velocity. The dynamic behavior of the particle is widely divergent from the particle in PSO as the position and velocity of the particles cannot be determined simultaneously. Only the probability of a particle appearing in a particular location $\overrightarrow{X}$ can be determined from the probability density function $|{\psi(\overrightarrow{x},t)}|^2$. A delta potential well is employed to constrain the quantum particles and prevent explosion. Since the search space and the solution space are different, a state transformation from the quantum state to classical state called 'collapse' is employed.\\
The proposed changes to the QPSO algorithm are related to the initialization of the particles as well as the position update rule for the algorithm. A chaotic initialization of the particles is done using the Lorenz attractor, which is a set of chaotic solutions for the Lorenz equation. The particle position update rule is changed to something similar to a Levy Flight mechanism, which is exhibited by animals when searching for food in an area.\\
The multi-objective problem that the algorithm will be fine tuned to optimize is the bi-objective Cobb Douglas Habitability function, which is used to generate the Cobb Douglas Habitability Score for exoplanets. The score is composed of two parts, namely the interior score and the surface score of the particular planet.
\section{Cobb Douglas Habitability Function}
 The general motivation for using Cobb-Douglas production function is because of its interesting properties. It is a function that models the response of an output parameter on varying its inputs. The function is concave when the sum of elasticities is not greater than one ensuring that an optimum exists which maximizes the function inside a feasible region defined by the constraints on elasticities . It was first originally introduced to model the growth of American economy during 1899-1922. In the case of exoplanetary habitability, the proposed metric models how the habitability score $Y$ changes on varying input planetary parameters. This is achieved by allowing the coefficients of elasticity to be adjusted via an optimization algorithm. It has already been established that the proposed habitability metric consists of two components: surface score and interior score. The final CDHS, defined in equation , is equal to the convex combination of $Y_i$ and $Y_s$. The weights $\omega_i$ and $\omega_s$ defines the importance of interior score and surface score in determining the final CDHS, respectively. Here, $\omega_i$ and $\omega_s$ sum up to 1. Finally, the Cobb-Douglas Habitability production function can be formally written as
 \begin{equation}
     Y = R^\alpha . D^\beta . V_{e}^{\delta} . T_{s}^{\gamma}\label{eq1} 
 \end{equation}
 where $R$, $D$, $V_e$ and $T_s$ is the radius, density, escape velocity and surface temperature respectively. $\alpha$,$\beta$,$\delta$ and $\gamma$ are coefficients of elasticity and $ 0 < \alpha , \beta ,\gamma , \delta < 1$. 
\\
The Cobb Douglas Habitability score is estimated by breaking it up into the interior score ($CDHS_i$) and the surface score ($CDHS_s$) and maximizing the following production functions.
\begin{subequations}
\label{eq2}
\begin{equation}
Y_i = CDHS_i = R^\alpha . D^\beta \label{subeq1}
\end{equation}
\begin{equation}
    Y_s = CDHS_s = V_e^\gamma . T_s^\delta \label{subeq2}
\end{equation}
\end{subequations}
Equations \eqref{subeq1} and \eqref{subeq2} are convex under either Constant Returns to Scale (CRS), when $\alpha + \beta = 1$ and $\gamma +\delta = 1$, or Decreasing Returns to Scale (DRS), when $\alpha + \beta < 1$ and $\gamma + \delta < 1$. The final Cobb Douglas Habitability Score is the convex combination of the individual interior and surface scores, given by, 
\begin{equation}
    Y = \omega_i . Y_i + \omega_s . Y_s \label{eq3}
\end{equation}
+\section{Quantum-behaved Particle Swarm Optimization}
Quantum-behaved Particle Swarm Optimization is an improved version of the biologically inspired metaheuristic algorithm known as Particle Swarm Optimization, which is used to find the global minima of a function. In PSO, particles move around and converge towards the globally optimal solution while losing kinetic energy as they approach the solution, similar to how a particle would behave in a potential field of attraction at the optimal point. QPSO builds upon this by making use of quantum potential fields, and introducing the particles as quantum particles represented by their waveforms. Making use of a potential model, we can simulate the similar behaviour of particles being attracted to the centre of the quantum potential field. In most cases, the Delta Potential Well model is used for QPSO as it provides faster convergence, and this paper employs the same as introduced in \cite{QPSO-original}. 
\subsection{Proposed changes to the QPSO algorithm}
\subsubsection{Chaotic Initialization}
Chaos theory is a part of mathematics that looks at systems that are very sensitive. A very small change can make the system behave very differently. It deals with nonlinear things which are impossible to predict or control, like weather, turbulence, stock market etc. It is popularly known by the butterfly effect, in which the flapping of a butterfly's wings could lead to a hurricane somewhere else. It may take a long time to become a hurricane, but the connection still exists. Since the weather is a very sensitive system, the flapping of the wings at that point in space-time or a different time would have drastically different effects. This is the a simple example of a small change in the initial conditions leading to drastic changes over time.\\
Edward Lorenz, who was a mathematician and meteorologist, also known as the founder of modern Chaos Theory made a weather model which involved 12 differential equations and exhibited chaotic behavior. In his effort to find chaotic systems in simpler set of equations, he was led to the phenomenon of rolling fluid convection and came up with the following equations. 
\begin{subequations}
\begin{equation}
    \frac{dx}{dt} = \sigma(y-x) \label{lorenz1}
\end{equation}
\begin{equation}
    \frac{dy}{dt} = x(\rho - z) - y \label{lorenz2}
\end{equation}
\begin{equation}
    \frac{dz}{dt} = xy -\beta z \label{lorenz3}
\end{equation}
\end{subequations}

When the parameters of the system, $\sigma$ , $\rho$ and $\beta$ are $10$ , $28$ and $\frac{8}{3}$ respectively, the system described by Equations \eqref{lorenz1}, \eqref{lorenz2} and \eqref{lorenz3} displays chaotic behavior. 
\begin{figure}[htbp]
\centerline{\includegraphics[width=0.5\textwidth]{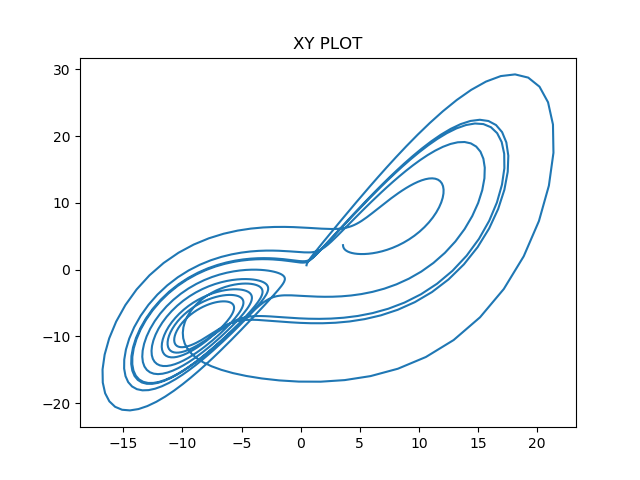}}
\caption{Projection of the Lorenz Chaos system on the XY plane}
\label{chaos_xy}
\end{figure}
\begin{figure}[htbp]
\centerline{\includegraphics[width=0.5\textwidth]{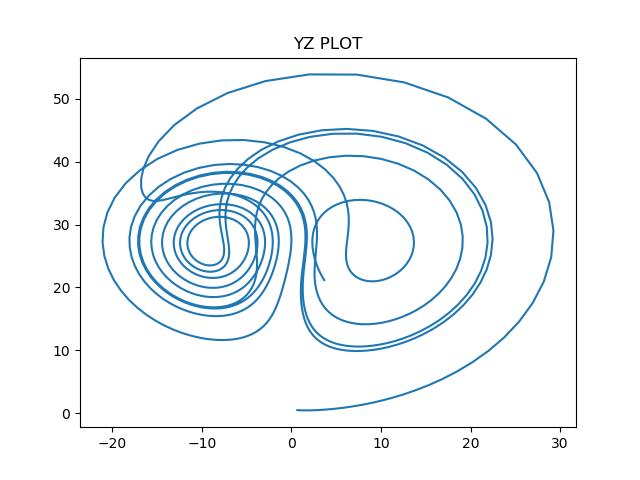}}
\caption{Projection of the Lorenz Chaos system on the YZ plane}
\label{chaos_yz}
\end{figure}
\begin{figure}[htbp]
\centerline{\includegraphics[width=0.5\textwidth]{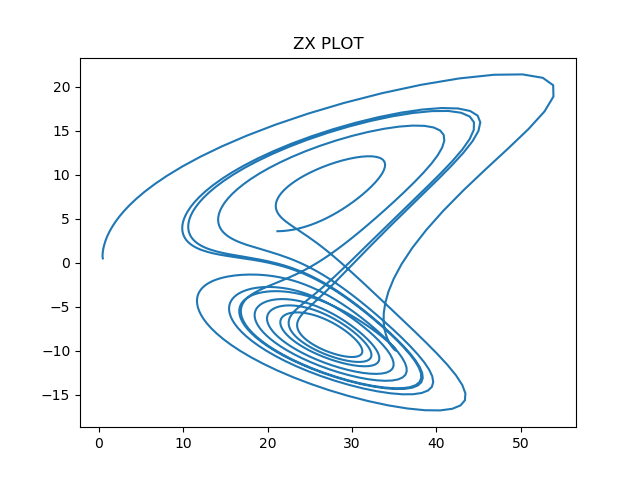}}
\caption{Projection of the Lorenz Chaos system on the ZX plane}
\label{chaos_zx}
\end{figure}
\\
From figures \ref{chaos_xy},\ref{chaos_yz} and \ref{chaos_zx} we can see that the Lorenz system of equations is a strange attractor. Since it is a chaotic strange attractor, which exhibits sensitive dependence on initial conditions, any two arbitrarily close alternative initial points on the attractor, after any of various numbers of iterations, will lead to points that are arbitrarily far apart (subject to the confines of the attractor), and after any of various other numbers of iterations will lead to points that are arbitrarily close together. Thus a dynamic system with a chaotic attractor is locally unstable yet globally stable: once some sequences have entered the attractor, nearby points diverge from one another but never depart from the attractor.\\
This behavior of the Lorenz system can be used to initialize particles in the Quantum-behaved Particle Swarm Optimization algorithm. A similar approach was followed in \cite{Chaotic_PSO}, where the CPSO algorithm used the Henon map and Tent Map as the chaotic initializations of the particles. In a similar way, the Lorenz system of equations is used to create a map which initializes all the particles in the modified QPSO algorithm. Since the Lorenz system is restricted to three dimensions, multiple Lorenz systems with different initializations are used for particles with higher dimensions. The dimensions are then scaled and mapped to the particles using the limits of the Lorenz system. The main objective behind the chaotic initialization is the importance of the initial positions of the particles as they can help resolve premature convergence which hinders the algorithm from finding the global minima of a given objective function. \\
\subsubsection{Levy Flight}
Levy Flight is a random walk in which the step-lengths have a probability distribution that is heavy-tailed. When defined as a walk in a space of dimension greater than one, the steps made are in isotropic random directions. Levy flight stems from the mathematics related to chaos theory and is useful in stochastic measurement and simulations for random or pseudo-random natural phenomena. Examples include earthquake data analysis, financial mathematics, cryptography, signals analysis as well as many applications in astronomy, biology, and physics. 
For general distributions of the step-size, satisfying the power-like condition, the distance from the origin of the random walk tends, after a large number of steps, to a stable distribution due to the generalized central limit theorem, enabling many processes to be modeled using Lévy flights.
The probability densities for particles undergoing a Levy flight can be modeled using a generalized version of the Fokker–Planck equation, which is usually used to model Brownian motion. The equation requires the use of fractional derivatives. For jump lengths which have a symmetric probability distribution, the equation takes a simple form in terms of the Riesz fractional derivative. In one dimension, the equation reads as,
\begin{equation}
    \frac{\delta \phi (x,t)}{\delta t} = -\frac{\delta}{\delta x} f(x,t) \phi (x,t) + \gamma \frac{\delta^\alpha\phi (x,t)}{\delta |{x}|^\alpha}
\end{equation}
where $\gamma$ is a constant akin to the diffusion constant, $\alpha$ is the stability parameter and f(x,t) is the potential. The Riesz derivative can be understood in terms of its Fourier Transform.
\begin{equation}
    F\left[\frac{\delta^\alpha\phi (x,t)}{\delta |{x}|^\alpha}\right] = k^\alpha F_k\left[\phi (x,t)\right]
\end{equation}
This naturalistic form of movement can be compared to organisms wandering away from regions of over-saturation, which in case of optimization problems is highly beneficial in allowing the model to explore a larger region in the solution space before complete convergence. The main objective in using Levy Flight in the QPSO model is that it is possible to simulate the wandering of particles away from global or known optima and improve the search abilities of the model for problems which have a high number of local optima, leading to greater frequency of convergence to the global optima.
\begin{figure}[htbp]
\centerline{\includegraphics[width=0.5\textwidth]{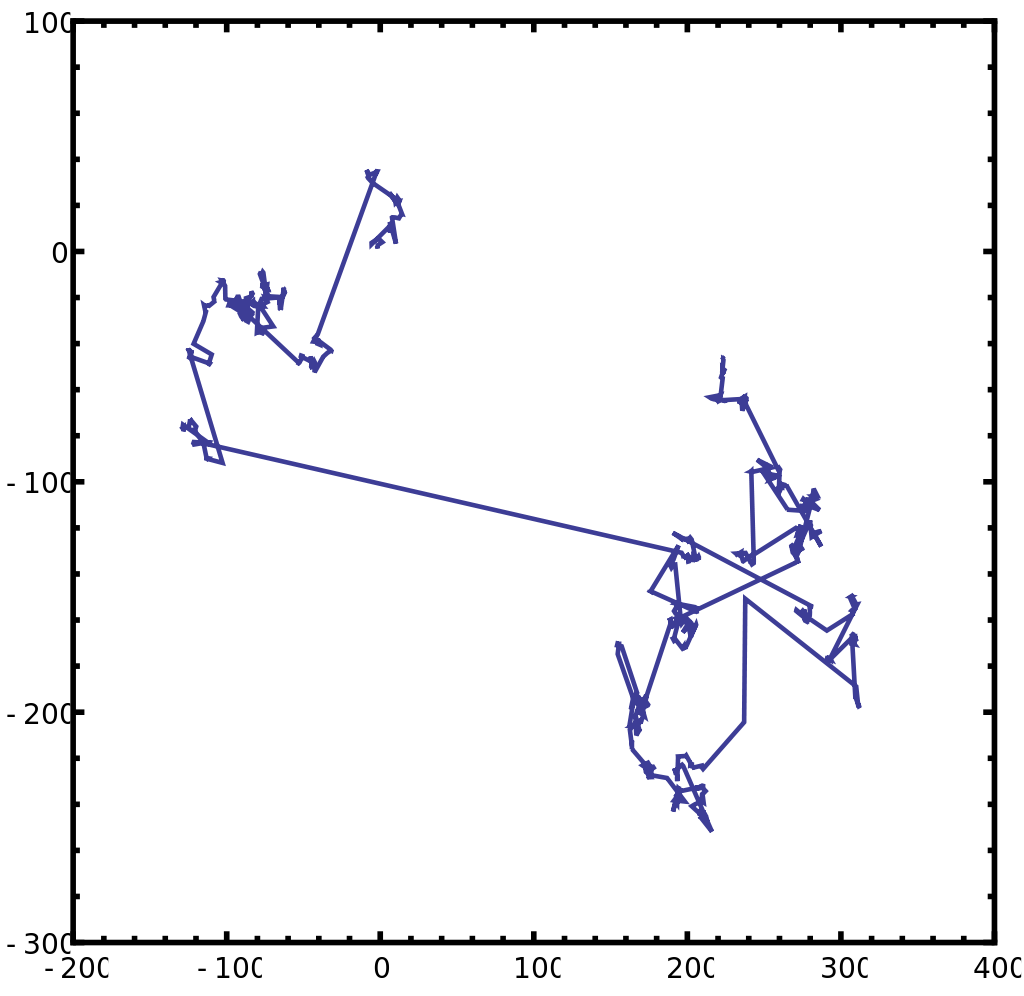}}
\caption{An example of 1000 steps of a Lévy flight in two dimensions}
\label{levy}
\end{figure}
\begin{figure}[htbp]
\centerline{\includegraphics[width=0.5\textwidth]{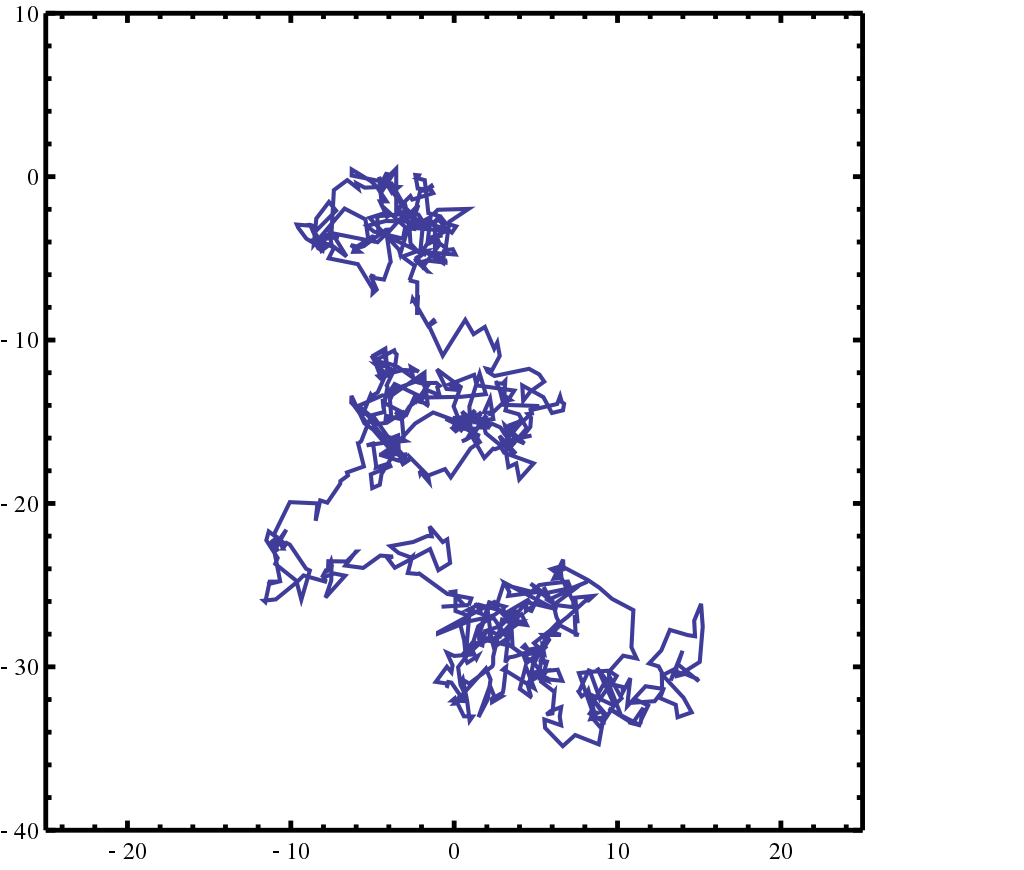}}
\caption{ An example of 1000 steps of an approximation to a Brownian motion type of Lévy flight in two dimensions}
\label{levy_b}
\end{figure}
\section{Representing the problem}
A Constrained Optimization problem can be represented as,
\begin{mini*}{x}{f(x)}{}{}
\addConstraint{g_k(x)}{\leq0 ,}{k=0 \ldots N-1}
\addConstraint{h_l(x)}{=0 ,\quad}{l=1 \ldots r}
\end{mini*}
Ray and Liew \cite{RayLiew} describe a way to represent non strict inequality constraints when optimizing using a particle swarm. Strict inequalities and equality constraints are to be converted to non strict equalities before representing them in the problem. Introducing an error threshold $\epsilon$ converts strict inequalities of the form $g_{k}'(x) < 0$ to non strict inequalities of the form $g_k(x) = g_{k}'(x) + \epsilon \leq 0$. A tolerance $\tau$ is used to convert equality constraints into a pair of inequalities,

\begin{equation}
\begin{matrix}
\displaystyle &g_{(q+l)}(x) &= &h_l(x) - \tau \leq 0, &l = 1...r,  \\
&g_{(q+r+l)}(x) &= &-h_l(x) - \tau \leq 0, &l = 1...r.
\end{matrix}
\end{equation}
In the above way, r equality constraints become 2r equality constraints, raising the total number of constraints to $s= q + 2r$. For each potential solution $p_i$, $c_i$ represents the constraint vector where, $c_{ik} = \max\{g_k(p_i),0\}$, $k = 1...s$. When $c_{ik}=0$, $\forall k = 1...s$, the solution lies within the feasible region. When $c_{ik}$ > 0, the potential solution $p_i$ violates the $k^{th}$ constraint.
\subsection{Representing CDH Score Estimation}
Similar to the way the CDH Score is represented in \cite{CEESA-PSO}, the CDH Score estimation under CRS is represented as,\\
\begin{mini!}{\alpha,\beta,\gamma,\delta}{Y_i=R^\alpha . D^\beta}{\label{crs}}{}
\breakObjective{Y_s = V_e^\gamma . T_s^\delta}
\addConstraint{-\phi+\epsilon}{\leq0,\label{crs:cons1}}{\forall\phi\in\{\alpha,\beta,\gamma,\delta\}}
\addConstraint{\phi-1+\epsilon}{\leq0,\label{crs:cons2}}{\forall\phi\in\{\alpha,\beta,\gamma,\delta\}}
\addConstraint{\alpha+\beta-1}{\leq0 \label{crs:cons3}}
\addConstraint{1-\alpha-\beta}{\leq0 \label{crs:cons4}}
\addConstraint{\gamma+\delta-1}{\leq0 \label{crs:cons5}}
\addConstraint{1-\gamma-\delta}{\leq0. \quad \label{crs:cons6}}{}
\end{mini!}
Under DRS, the constraints \eqref{crs:cons3} to \eqref{crs:cons6} are replaced with,
\begin{subequations}
\begin{equation}
\alpha + \beta + \epsilon - 1 \leq 0 \label{drs:cons1}
\end{equation}
\begin{equation}
\gamma + \delta + \epsilon -1 \leq 0 \label{drs:cons2}
\end{equation}
\end{subequations}
\section{Experiments and Results}
\subsection{Testing the Proposed Algorithm}
The proposed changes in the algorithm are first tested on a series of test functions given by
\begin{itemize}
    \item Rosenbrock Function: $f(x)=(1-x)^2 + 100(y-x^2)^2$
    \item Mishra Bird Function: $f(x,y)=e^{(1-\cos{y})^2}\sin{x} + e^{(1-\sin{x})^2}\cos{y} + (x-y)^2$
    \item Ackley Function: $f(x.y) = -20e^{0.2^2 \sqrt{(x^2+y^2})} + e^{0.5+ \cos{2\pi x} + \cos{2\pi y}} + e + 20$
    \item Levi function: $f(x,y) = \sin^2{(3\pi x)} + (x-1)^2(1 + \sin^2{(3\pi y)}) + (y-1)^2(1+\sin^2{(2\pi y)})$
\end{itemize}
The algorithm was named LQPSO and another variant of it was named as the LDQPSO. The LDQSP algorithm added a Levy Flight decay, which decayed the effect of the Levy flight as the number of iterations increased. For all the algorithms, the total number of particles were $1000$ and the LQPSO and the LDQPSO algorithms were initialized using Lorenz Chaos Map. The Lorenz Map is a similar one as used in \cite{Chaotic_PSO}. Each of the algorithms were tested a total of $30$ times to get their average iterations and to calculate their success rate. The results are presented in Table \ref{tab2}.

\begin{algorithm}
    \caption{LDQPSO minimization}\label{alg}
    \begin{algorithmic}[1]
        \Procedure{minimize}{$fun,x$}\Comment{Minimizing function fun using initialised particles x0}
        \Repeat
            \State $pbest\gets x$
            \State $gbest =$ getBest($fun, pbest$)\Comment{Best solution in $pbest$ for $fun$}
			\For{$i\gets 1$ to $population size$ $M$}
				\If{fun($x_i$) $<$ fun($pbest_i$)}
						\State $pbest_i = x_i$
				\EndIf
				\State $u =$ rand($0,1$)
				\State $f_1 =$ rand($0,1$),$f_2 =$ rand($0,1$)
				\State $P = (f_1*gbest + f_2*pbest)/(f_1+f_2)$
				\State find $mbest$
				\For{$d\gets 1$ to $dimension$ $D$}
					\State $l = DecayedLevyWalkFactor()$
					\State $update = mbest_d*l*$ln($1/u_d$)
					\If{random($0,1$)$ > 0.5$}
						\State $position_d = P_d - update$
					\Else
						\State $position_d = P_d + update$
					\EndIf
				\EndFor
				\State $gbest =$ getBest($fun, pbest$)
			\EndFor
		\Until{termination criteria is met}
		\EndProcedure
	\end{algorithmic}
\end{algorithm}

\begin{table}[htbp]
\caption{Avg iterations to convergence}
\begin{center}
\begin{tabular}{|c|c|c|c|c|}
\hline
\textbf{\textit{Test}}&\multicolumn{4}{|c|}{\textbf{Algorithms}} \\
\cline{2-5} 
\ \textbf{\textit{Functions}} &\textbf{\textit{PSO}}&\textbf{\textit{QPSO}}&\textbf{\textit{LQPSO}}&\textbf{\textit{LDQPSO}} \\
\hline
Ackley& $194$ & $32.6$ & $33.834$ & $45.2$\\
\hline
Levi& $130.934$ & $46.734$ & $48.734$ & $49.734$\\
\hline
Rosen& $134.767$ & $42.8$ & $42.934$ & $46.8$\\
\hline 
Mishra Bird &$152.734$ & $77.534$ & $65.634$ & $60.534$\\
\hline
\end{tabular}
\label{tab1}
\end{center}
\end{table}

The results presented in Table \ref{tab1} were generated using the Psopy library which was created as part of the paper in \cite{CEESA-PSO}. All the algorithms had a $100\%$ success rate on all the test functions except for the Mishra Bird function. PSO had the lowest success rate of $~83\%$, with QPSO having a better success rate of $90\%$, with the LQPSO and the LDQPSO both having a success rate of $93\%$. This shows that the proposed algorithm does better at avoiding the local optima compared to the original PSO as well as the revised QPSO algorithm. 
\subsection{Testing the Algorithm on the CD-HPF}
After seeing the results of the LDQPSO algorithm on the test functions, the LDQPSO algorithm was used to optimize the Cobb Douglas Habitability function using a modified version of the jMetalPy framework \cite{jMetalPy} . A subset of the original PHL-EC Dataset \cite{PHLEC} was used, specifically the exoplanets belonging to the TRAPPIST-1 system. The Pareto front plots were generated using $25$ particles just as in \cite{CEESA-PSO}.  \\
\begin{figure}[htbp]
\centerline{\includegraphics[width=0.5\textwidth]{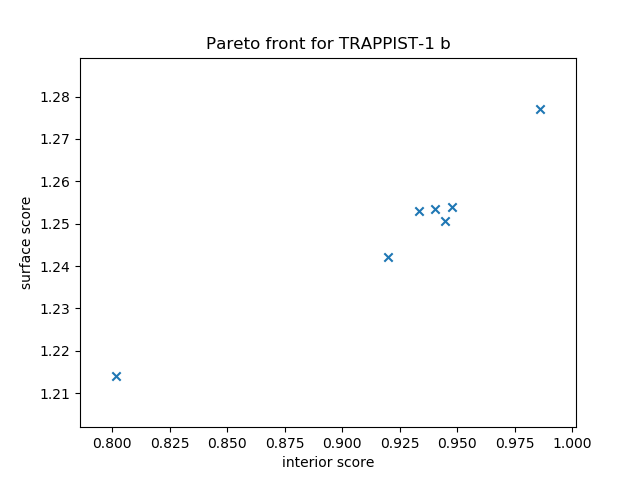}}
\caption{Trappist-1b under CRS conditions}
\label{Trappist1bCRS}
\end{figure}
\begin{figure}[htbp]
\centerline{\includegraphics[width=0.5\textwidth]{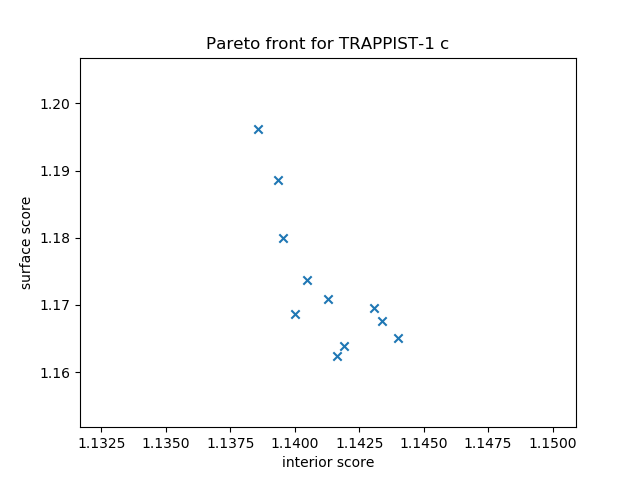}}
\caption{Trappist-1c under CRS conditions}
\label{Trappist1cCRS}
\end{figure}
\begin{figure}[htbp]
\centerline{\includegraphics[width=0.5\textwidth]{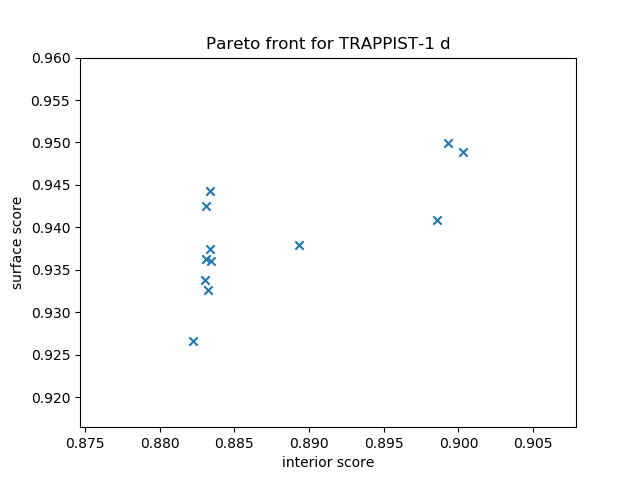}}
\caption{Trappist-1d under CRS conditions}
\label{Trappist1dCRS}
\end{figure}
\begin{figure}[htbp]
\centerline{\includegraphics[width=0.5\textwidth]{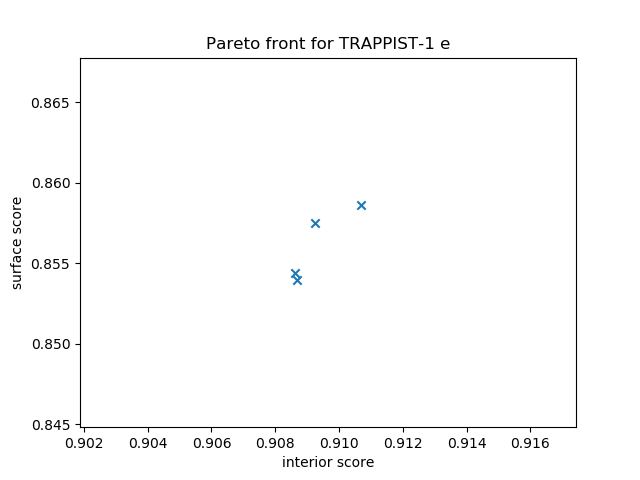}}
\caption{Trappist-1e under CRS conditions}
\label{Trappist1eCRS}
\end{figure}
\begin{figure}[htbp]
\centerline{\includegraphics[width=0.5\textwidth]{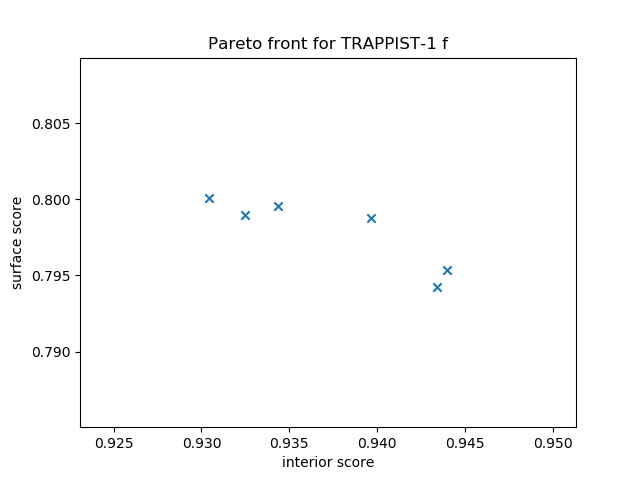}}
\caption{Trappist-1f under CRS conditions}
\label{Trappist1fCRS}
\end{figure}
\begin{figure}[htbp]
\centerline{\includegraphics[width=0.5\textwidth]{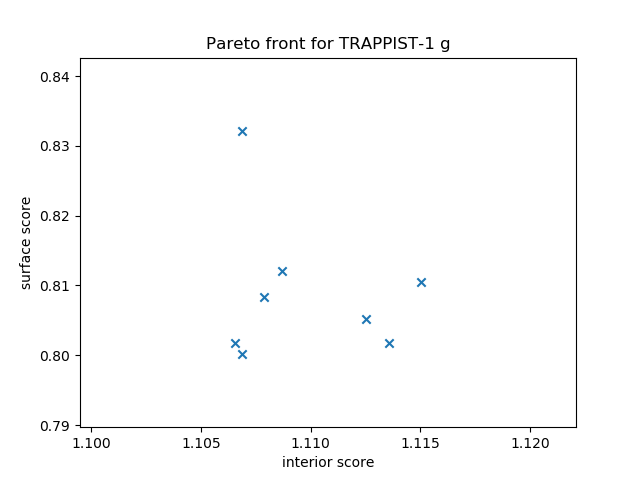}}
\caption{Trappist-1g under CRS conditions}
\label{Trappist1gCRS}
\end{figure}
\begin{figure}[htbp]
\centerline{\includegraphics[width=0.5\textwidth]{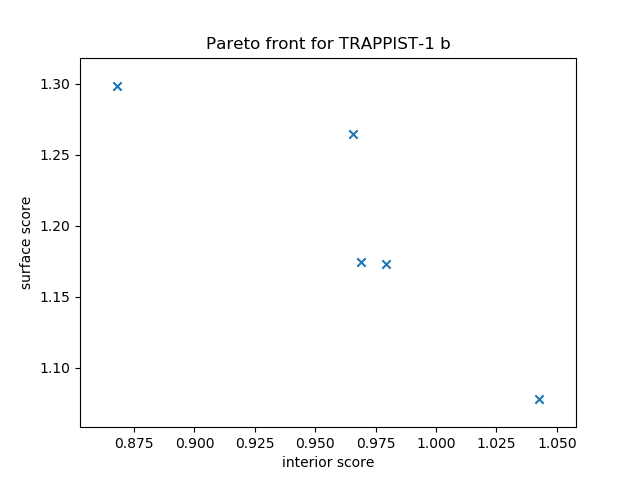}}
\caption{Trappist-1b under DRS conditions}
\label{Trappist1bDRS}
\end{figure}
\begin{figure}[htbp]
\centerline{\includegraphics[width=0.5\textwidth]{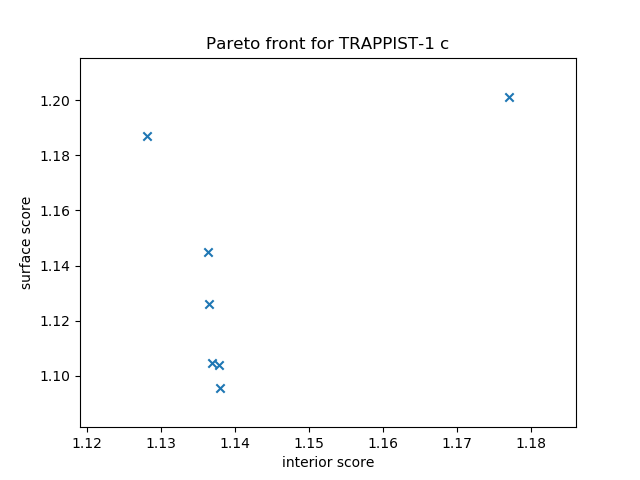}}
\caption{Trappist-1c under DRS conditions}
\label{Trappist1cDRS}
\end{figure}
\begin{figure}[htbp]
\centerline{\includegraphics[width=0.5\textwidth]{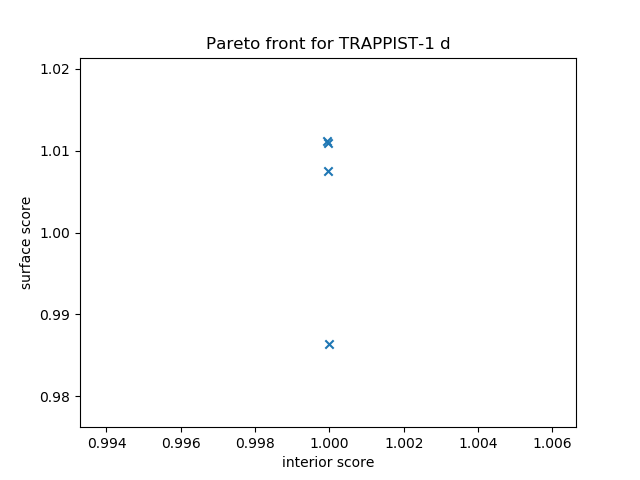}}
\caption{Trappist-1d under DRS conditions}
\label{Trappist1dDRS}
\end{figure}
\begin{figure}[htbp]
\centerline{\includegraphics[width=0.5\textwidth]{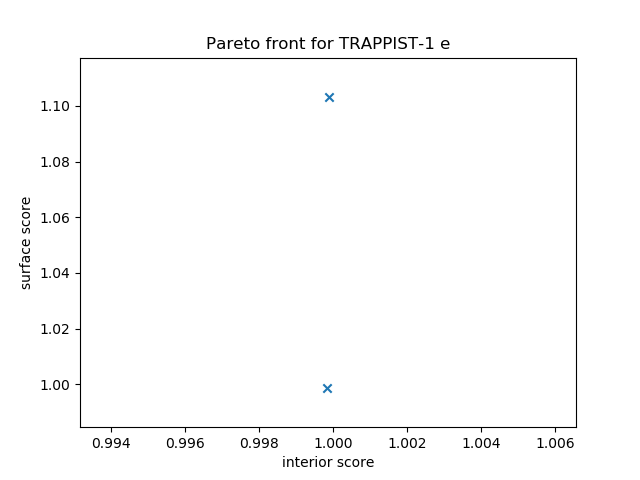}}
\caption{Trappist-1e under DRS conditions}
\label{Trappist1eDRS}
\end{figure}
\begin{figure}[htbp]
\centerline{\includegraphics[width=0.5\textwidth]{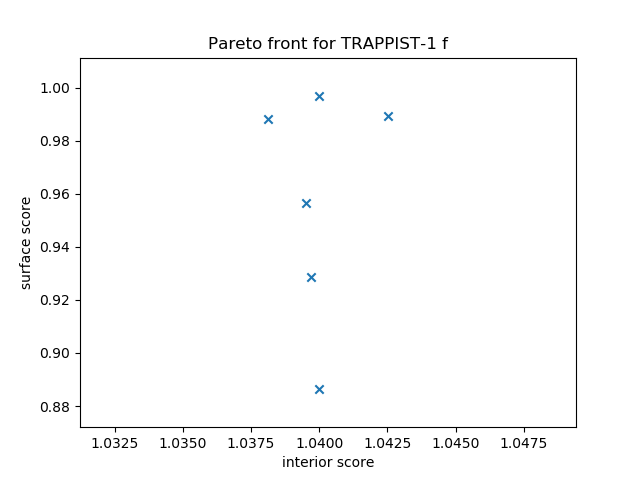}}
\caption{Trappist-1f under DRS conditions}
\label{Trappist1fDRS}
\end{figure}
\begin{figure}[htbp]
\centerline{\includegraphics[width=0.5\textwidth]{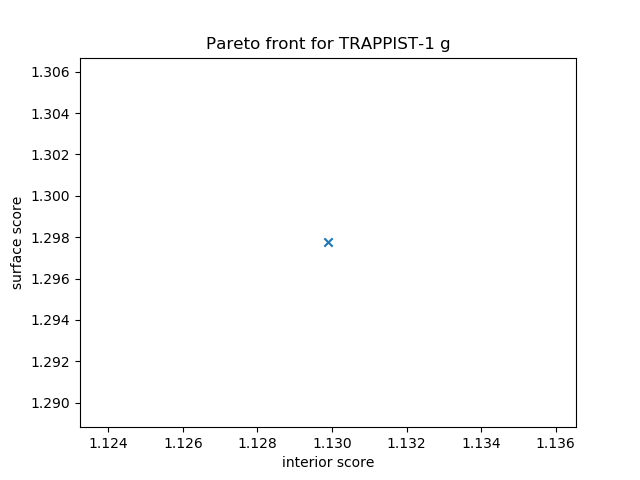}}
\caption{Trappist-1g under DRS conditions}
\label{Trappist1gDRS}
\end{figure}
\\
After comparing the results of the algorithm with the catalog mentioned in \cite{CEESA-PSO}\cite{CD-HPF} , we noticed that in most cases the points on the plot showed better scores than those mentioned in the catalog. In the original paper \cite{CD-HPF} however, the plots showed that most of the points were close together and a proper front could be seen. We think that this is not visible in the Pareto front plots generated by us as the conflict between the objective functions (for calculating CDHS) might be minimal and not visible in these plots as the convergence may be faster. Table \ref{tab2} shows the average number of iterations over $30$ runs that the LDQPSO algorithm took to converge for both Constant Returns to Scale as well as Decreasing Returns to Scale. A point to note is that unlike the offset in \cite{CEESA-PSO} , there is no offset in LDQPSO, due to which it may look misleading that the LDQPSO takes more iterations than PSO, but that is not the case as there is no offset used. We can clearly see that LDQPSO algorithm requires lesser number of iterations to converge.
\begin{table*}[ht]
\caption{Avg iterations to convergence}
\begin{center}
\begin{tabular}{|c|c|c|c|c|c|c|}
\hline
\textbf{\textit{Returns}}&\multicolumn{6}{|c|}{\textbf{Planets}} \\
\cline{2-7} 
\ \textbf{\textit{to Scale}} &\textbf{\textit{Trappist-1 B}}&\textbf{\textit{Trappist 1 C}}&\textbf{\textit{Trappist 1 D}}&\textbf{\textit{Trappist 1 E}} &\textbf{\textit{Trappist 1 F}} &\textbf{\textit{Trappist 1 G}} \\
\hline
CRS& $28.3$ & $26.067$ & $29.567$ & $28.034$ &$28.3$ &$26.734$\\
\hline
DRS& $52.3$ & $65.034$ & $37.034$ & $32.834$ &$73.767$ &$90.3$\\
\hline
\end{tabular}
\label{tab2}
\end{center}
\end{table*}

\begin{table*}[ht]
\caption{CDHS Scores}
\begin{center}
\begin{tabular}{|c|c|c|c|c|c|c|}
\hline
\textbf{\textit{Conditions}}&\multicolumn{6}{|c|}{\textbf{Planets}} \\
\cline{2-7} 
\ &\textbf{\textit{Trappist-1 B}}&\textbf{\textit{Trappist 1 C}}&\textbf{\textit{Trappist 1 D}}&\textbf{\textit{Trappist 1 E}} &\textbf{\textit{Trappist 1 G}} &\textbf{\textit{Trappist 1 H}} \\
\hline
CRS (0.5i + 0.5s)& $1.1929$ & $1.114$ & $0.7926$ & $0.887$ &$1.065$ &$0.7382$\\
\hline
DRS (0.5i + 0.5s)& $1.1887$	&$1.1717$	&$1.009$	&$0.9976$	&$1.0795$	&$0.9645$\\
\hline
CRS (0.6i + 0.4s)& $1.1612$	&$1.132$	&$0.8005$	&$0.888$	&$1.0730$	&$0.7452$\\
\hline
DRS (0.6i + 0.4s)& $1.1619$	&$1.1675$	&$1.0073$	&$0.9977$	&$1.0803$	&$0.9695$\\

\hline
\end{tabular}
\label{tab3}
\end{center}
\end{table*}
\section{Conclusion}
Quantum-behaved Particle Swarm Optimization as a variant of the original Particle Swarm Optimization is a highly parallelizable and easy to implement algorithm, which performs better than the original PSO proposed by Kennedy and Ebenhart in \cite{PSO_Org} . Since it does not need any gradient calculation, it can work in high dimensional search spaces with a large number of constraints, which is useful in cases such as a Habitability score estimate where many input parameters can be used. The particles in QPSO are independent of each other in a single iteration, allowing their updates to happen simultaneously and asynchronously. \\
Although the results of the Quantum-behaved Particle Swarm Optimization and Particle Swarm Optimization algorithms are not as accurate as direct methods, the scaling of the algorithms when the number of input parameters increases allows it to be more feasible than traditional optimization methods as it can handle the higher number of constraints. \\
The main aim of this manuscript is to compare the performance of the Quantum-behaved Particle Swarm Optimization with the Particle Swarm Optimization while proposing some changes to the model itself. These changes are influenced by Chaos Theory and the movement of animals foraging for food in an area. As we observed from our experiments, the modified QPSO algorithm performed better than the PSO in terms of performance and it's ability to avoid getting stuck in a local optima. 
\section{APPENDIX}
\subsection{Gradient Simulation of QPSO}
The QPSO system functions by initializing a set of particles each with a pseudo-random wave function. The position of the particles as obtained from these wave functions at each iteration describe each particle's solution at that instance, which are feasible at initialization. However, the position of the particle is updated on every iteration of the process which might put the particle on an infeasible solution. Now, we simulate a particle well for each particle such that each particle well has a centre at point p, which is related to the wave function of the particle by,\\

\begin{equation}
    d^2\psi/dy^2=2m/h [E+\gamma\delta(y)]\psi
\end{equation}

Hence, at each iteration, the algorithm stores a set of feasible solutions L represented by the the position of each particle, pi as well as the globally optimal solution gbest represented by pg. At the start of the process, the algorithm initializes L to the initial positions of the particles and gbest to the best solution in L. At each iteration, QPSO calculates the position of each particle at that instance using their p value and the gbest value pg by simulating the delta potential well with a characteristic length l determined by the gbest as,\\

\begin{equation}
    L = 1/\beta = \hbar/(m\gamma)
\end{equation}

The algorithm then silumates a gradient based on the random new position of the particle at the instance using the current delta potential well, which will push it either towards or away from the gbest value. This new position is then used as the new centre of each particle's respective delta potential well. Each iteration can hence be summed up as,\\

\begin{equation}
    x=P \pm L/2 ln(1/u)
\end{equation}

where x is the new solution obtained, u is a uniform random number,  and the movement is simulated by the obtained position in the delta potential well of each particle, where each particle's p will move with larger steps towards or away from the current solution based on the characteristic length of the well at that point given by l.
However, there remains a probability that the new solution may not have been feasible or that it may have been less optimal than the prior position's solution due to the random nature of the obtained position at that instance. Hence, we shall add the rule,
\begin{equation}
\text{if} f(x_i )<f(p), \hspace{0.2mm} \text{then} \hspace{0.2mm} p=x_i
\end{equation}

which guarantees that there will be convergence and that the particles do not move away from their optimal and opposite to the gradient. Here, the update of the centre of the delta potential well for each particle is analogous to the update of its wave function as the two are directly related. Once the positions are updated, the algorithm then updates L and gbest as discussed earlier.
After each iteration, each particle moves a little closer toward gbest while the particle at gbest also moves and possibly finds a better solution. This in turn leads to L and gbest being updated in case any of the particles come across better solutions. Eventually after several iterations, the particles and their corresponding pi values will converge toward a gbest solution.\\
\section{ACKNOWLEDGMENT}
We would like to thank the Department of Computer Science and Engineering at PES University, for encouraging and supporting us in writing this manuscript. We would also like to thank Dr. Snehanshu Saha for laying the groundwork through his endeavours in the field of Astro-Informatics and providing his expertise and knowledge during the course of the research.
\nocite{*}
\bibliographystyle{ieeetr}
\bibliography{sample}
\end{document}